\documentclass[aps,twocolumn,showpacs]{revtex4}
\usepackage{psfig}

\def\order#1{{\cal O}\left(#1\right)}
\newcommand{\ba}{\begin{eqnarray}}
\newcommand{\ea}{\end{eqnarray}}
\newcommand{\p}{\mbox{$\vec{p}$}}
\newcommand{\e}{\mbox{$\vec{e}$}}
\newcommand{\n}{\mbox{$\vec{n}$}}
\newcommand{\vsig}{\mbox{$\vec{\sigma}$}}
\newcommand{\be}{\begin{equation}}
\newcommand{\ee}{\end{equation}}
\def\eq#1{(\ref{#1})}

\begin{document}

\title{
%
%
\[ \vspace{-2cm} \]
\noindent\hfill\hbox{\rm  } \vskip 1pt
\noindent\hfill\hbox{\rm Alberta Thy 13-01} \vskip 1pt
\noindent\hfill\hbox{\rm SLAC-PUB-8972} \vskip 1pt
\noindent\hfill\hbox{\rm hep-ph/0108233} \vskip 10pt
Top Quark Threshold Production at a $\gamma\gamma$ Collider
at Next-to-Next-to-Leading Order
}

\author{Andrzej Czarnecki}
\affiliation{
Department of Physics, University of Alberta\\
Edmonton, AB\ \  T6G 2J1, Canada\\
E-mail:  czar@phys.ualberta.ca}

\author{Kirill Melnikov}
\affiliation{
Stanford Linear Accelerator Center\\
Stanford University, Stanford, CA 94309\\
E-mail: melnikov@slac.stanford.edu}

\begin{abstract}
The next-to-next-to-leading order (NNLO) QCD corrections  to top quark 
threshold pair production  in $\gamma \gamma$ collisions are computed.
The NNLO effects turn out to be significant.  
Threshold cross sections 
in $\gamma\gamma$ and $e^+e^-$ collisions are compared.
\end{abstract}


\maketitle

Studies of threshold production of top quark pairs will provide
information about Standard Model parameters and a window to
``New Physics'' phenomena.  Because the $t$ quark decays fast compared
to the strong interaction time scale, even the threshold top pair
production in $e^+e^-$ or $\gamma\gamma$ collisions can be described
using perturbative QCD \cite{Fadin:1987wz}.  However, close to the
threshold, the cross section is enhanced by the strong
attraction between the quark and the antiquark and an all-order
resummation of the Coulomb effects is required.  The magnitude of the
threshold cross section and the position of its peak are sensitive to
the top quark mass and width, as well as the strong
coupling constant $\alpha_s$, and can help determine those parameters
with high precision \cite{topphys,Hoang:2000yr}.  In addition,
precision measurements in the threshold region at the Next Linear
Collider (NLC) will probe  electroweak interactions of the $t$
quarks and be sensitive to ``New Physics''.

A particularly interesting environment for studying ``New Physics''
effects in the top production is the $\gamma\gamma$ collider mode of
the NLC, with laser beams backscattered from the high-energy $e^+$ and
$e^-$ as the source of the colliding photons \cite{Bigi:1993pr}.  It
has been pointed out \cite{Hewett:1998ax,Boos:2000ki} that the process
$\gamma\gamma\to t\bar t$ is twice as sensitive to anomalous
photon-top quark couplings as $e^+e^- \to t\bar t$, and in addition is
free from contamination by the $Zt\bar t$ coupling.  It will therefore
allow searches for anomalous couplings of the top quark to photons, in
particular the electric dipole moment.  Although precision studies
might be feasible only at the NLC, the observation of $\gamma\gamma\to
t\bar t$ may be possible already at the LHC \cite{Piotrzkowski:2000rx}.

Although the threshold production cross section can be computed
perturbatively, it is not clear how well the series converges.
Recently, the renormalization group technique was employed
\cite{Luke:1999kz} to sum up large logarithmic corrections appearing
in  threshold problems.  It was applied to $\sigma(e^+e^- \to t
\bar t)$ at the threshold in \cite{Hoang:2001mm} and an
achievement of a 3\% accuracy in the normalization of the cross
section was claimed.  However, it is conceivable that possible large
non-leading-logarithmic effects may spoil this picture.  To a large
extent, study of the $t \bar t$ threshold production cross section in
the $\gamma \gamma$ mode might help in an assessment of the situation
and provide an additional test of the assumptions of the resummation
program.

In general, behavior of the $t\bar t$ threshold cross sections is
rather similar in $e^+e^-$ and $\gamma\gamma$ collisions; the
difference comes from the fact that the quarks are mainly produced in
spin one and spin zero state in $e^+e^-$ and $\gamma \gamma$
collisions, respectively.  Since the Coulomb interaction does not
depend on the spin, this fact does not affect the behavior of the
leading order cross section but, since relativistic corrections do
depend on the total spin of the produced fermion pair, it becomes
important in higher orders.

There is yet another interesting feature of the top quark threshold
production in $\gamma\gamma$ collisions \cite{Bigi:1993pr}.
In contrast to an $e^+e^-$ collider, manipulating
polarization of the incoming photons can suppress the  $S$-wave
production  and provide a possibility to
study the $P$-wave threshold production without the huge $S$-wave
background.  All these features of the top quark threshold 
in $\gamma \gamma$ collisions make it a very interesting laboratory
for studying the non-relativistic QCD dynamics of a heavy quark
anti-quark system, determining the standard model parameters, and
searching for ``New Physics''.

A known  problem with  this program is  the monochromaticity  of the
photon beams  at the NLC where  a typical energy  spread is currently
estimated to be about  10\% of the total energy 
(see e.g.~\cite{Abe:2001wn}). This may hamper the
threshold studies since a large energy spread will  wash 
out
any pronounced signal at the top  threshold.  It remains to be seen if
this difficulty can be overcome.

Much effort has already been spent on theoretical studies of
$\sigma(\gamma\gamma \to t\bar t)$ with polarized photons.  Away from
the threshold, one-loop QCD corrections were computed in
\cite{Kamal:1995ct}.  Close to the threshold, the cross section
$\sigma(e^+e^- \to t \bar t)$ has been studied in NLO in
\cite{Bigi:1993pr}.  Recently, studies of NNLO corrections to
$\sigma(e^+e^- \to t \bar t)$ were completed and the appropriate
effective field theory framework for studying higher order corrections
to the threshold phenomena was formulated (see \cite{Hoang:2000yr} and
references therein).  Applying similar techniques to $\gamma \gamma \to
t \bar t$, it is in principle straightforward to obtain the NNLO
corrections to this process; the major obstacle has been the lack of
the two-loop matching coefficient for an effective operator
responsible for the leading order transition $\gamma \gamma \to t\bar
t$ at the threshold. In this paper we describe the calculation of this
matching coefficient and present the NNLO corrections to
$\sigma(\gamma\gamma \to t\bar t)$ at the threshold.

Close to the threshold, the produced top quarks are non-relativistic 
so that it is reasonable to expand the production 
amplitudes in powers of the relative velocity $\beta$ of $t$ and $\bar 
t$. 
The Born amplitude for
$\gamma \gamma \to t \bar t$ with ${\cal O}(\beta^2)$ accuracy 
can be written as
\be
{\cal M} = -8i\pi\alpha Q_t^2 \phi^+ \left [ 
{\cal M}_S + {\cal M}_P \right ] \chi,
\ee
where $Q_t=2/3$ is the electric charge of the top quark, 
$\phi$ and $\chi$ are the top and anti-top spinors.
The amplitudes that give rise to the production of the $t \bar t$ 
state in $S$ and $P$ waves are, respectively
\ba
{\cal M}_S &=& \left (1 + \frac{(\p \n)^2}{m^2} -\frac{\p^2}{2m^2} 
\right ) 
i\n \cdot [\e_1 \times \e_2 ],
\\
{\cal M}_P&=& \frac{(\p \n) (\vsig \n) (\e_1 \e_2)+
(\p \e_1)(\vsig \e_2) + (\p \e_2)(\vsig \e_1)}{m}.
\nonumber 
\ea
In these formulas, $\n$ is a unit vector along the photon flight
direction, $\p$ is the top quark three-momentum and $\e_{1,2}$ are the
polarization vectors of the colliding photons.  From now on we will
consider the incoming photons to be circularly polarized. If the two
photons have opposite polarizations, the $S$ wave production amplitude
vanishes. As a consequence, only ${\cal M}_P$ contributes to the top
production and the cross section at the threshold behaves as
$\sigma_{+-} \sim \beta^3$. If the photon helicities are the same, the
top quarks are mainly produced in the $S$ wave with a small admixture
of the $P$ wave appearing in the NNLO.

It is customary to define a normalized unpolarized cross section,
$R_{\gamma \gamma} = \sigma(\gamma \gamma \to t \bar t)/\sigma_0$,
with $\sigma_0 = 4\pi \alpha^2/(3s), $ and decompose it into
contributions with the same and opposite photon helicities, $R_{\gamma
\gamma} = (R^{++}+ R^{+-})/2,$ where we have used $R^{++} = R^{--}$
and $ R^{+-}= R^{-+}$, valid for electromagnetic processes.  Let us
consider the case of the two photons with equal helicities and select
the contribution of the $S$ wave. (The case of the $P$ wave production
has been investigated earlier in the literature (for the most recent
discussion see \cite{Penin:1998ik}) and, because of its relative
${\cal O}(\beta^2)$ suppression, we do not have much to add to this
issue.) In case of the $S$ wave, the quarks are produced in a spin
singlet state. The expression for the cross section in the regime
$\alpha_s \ll \beta \ll 1$ reads:
\ba
R^{++}_S &=& 6 Q_t^4 N_c \beta 
\left ( 1 - \frac{\beta^2}{3} \right ) 
\nonumber \\
&& \cdot \left [ 1  + C_F \left ( \frac{\alpha_s}{\pi} \right ) 
\Delta^{(1)} 
+ C_F \left  ( \frac{\alpha_s}{\pi} \right )^2 \Delta^{(2)} \right],
\label{eq:rplusplus}
\ea
where $\beta = \sqrt{1-4m^2/s}$. Also, 
\be
\Delta^{(1)} = \frac{\pi^2}{2\beta} - \left ( 5- \frac{\pi^2}{4} \right 
) 
 + \frac{\pi^2}{2}\beta + \order{\beta^2}.
\ee
The two-loop corrections can be decomposed into four parts, arising
from the abelian and non-abelian gluon effects, and to light quark and
top quark vacuum polarization insertions in the one-loop correction.
Neglecting terms of $\order{\beta}$, the two-loop corrections are
\ba
\Delta^{(2)} =&&
\hspace*{-5mm}
 C_F \Delta_{A} + C_A \Delta_{NA} + T_R N_L \Delta_L + 
T_R N_H \Delta_H,
\\
\Delta_A &=& \frac{\pi^4}{12 \beta^2}
+ \left (- \frac{5}{2}+\frac{1}{8}\pi^2 \right ) \frac{\pi^2}{\beta}
\nonumber \\
&&+ \frac{27}{8}\pi^2+ \frac{25}{4}+ \frac{35}{192} \pi^4
- 2 \pi^2\ln( 2 \beta)+2 x_A;
\nonumber \\
 \Delta_{NA} &=& \left ( \frac{31}{72}
-\frac{11}{12} \ln(2 \beta) \right ) \frac{\pi^2}{\beta}
\nonumber \\
&& +\pi^2 \left ( \frac{5}{4} - \ln(2 \beta) \right ) 
+2 x_{NA};
\nonumber \\
 \Delta_{L} &=& \left ( -\frac{5}{18} + \frac{1}{3} 
\ln( 2 \beta) \right )\frac{\pi^2}{\beta} +2 x_L;
\nonumber \\
 \Delta_H &=& 2 x_H.
\ea
In the above formulas the terms $x_A,x_{NA},x_{L},x_{H}$ are related to 
the 
hard renormalization of the operator 
$\phi^+ \n \cdot \left [ \e_1 \times \e_2 \right ] \chi$
responsible for  $\gamma \gamma \to t \bar t$ transition at the 
threshold:
\ba
&& x_A = -21.02;~~~x_{NA} = -4.79;
\nonumber \\
&& x_L = -0.565;~~~x_H = 0.224.
\ea

\begin{figure}[htb]
\hspace*{-38mm}
\begin{minipage}{16.cm}
\begin{tabular}{ccc}
\psfig{figure=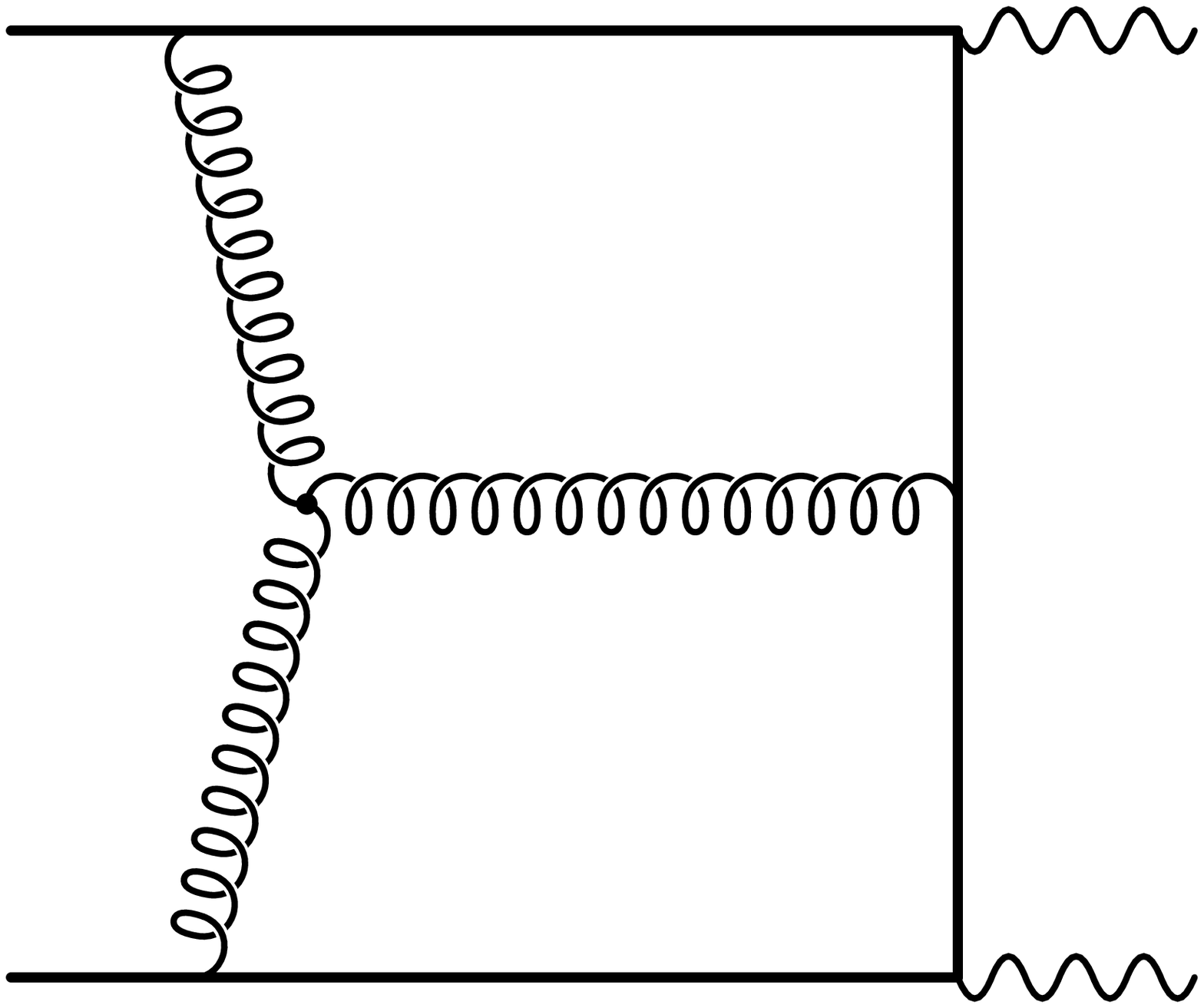,width=25mm}
&\hspace*{2mm}
\psfig{figure=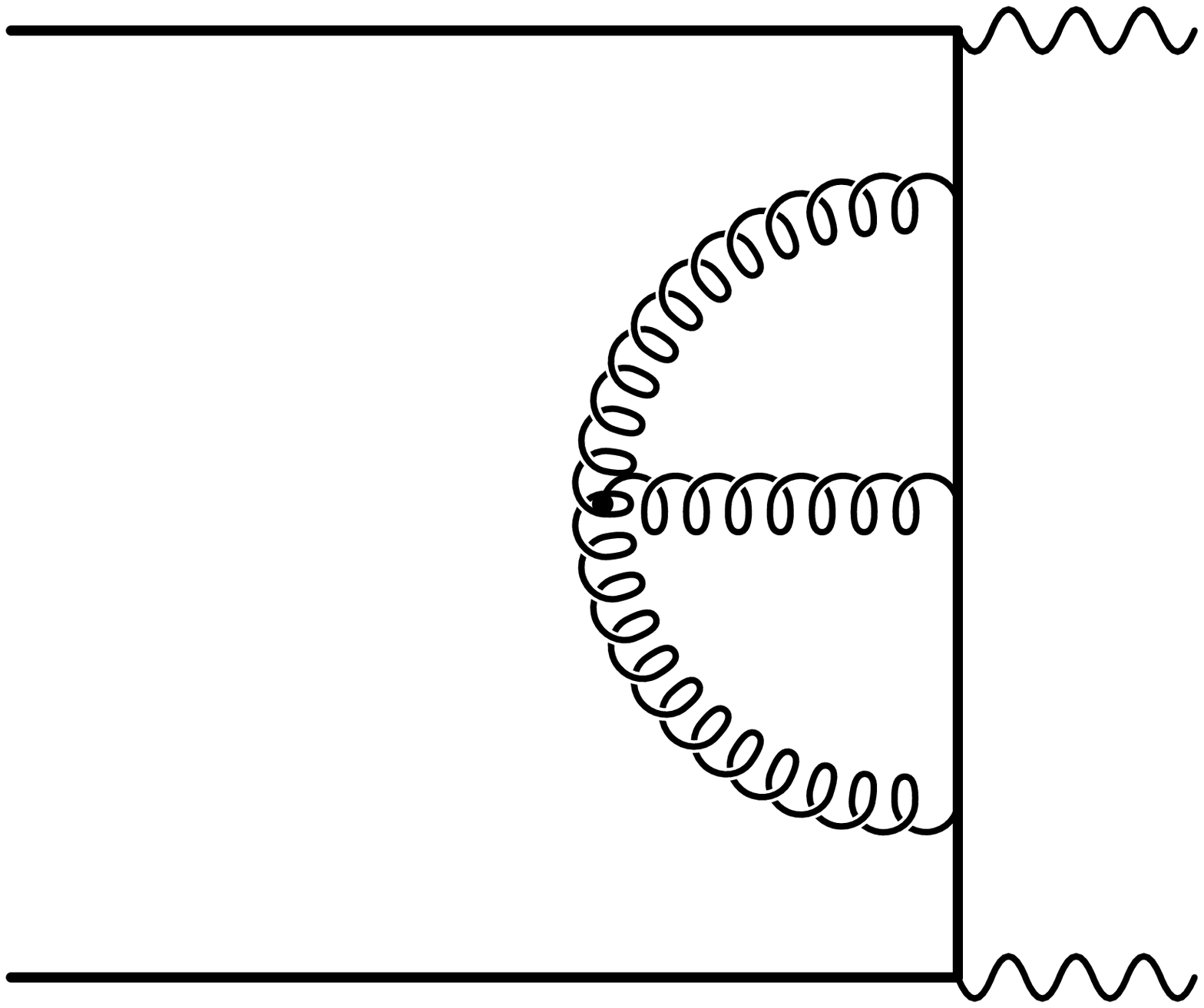,width=25mm}
&\hspace*{2mm}
\psfig{figure=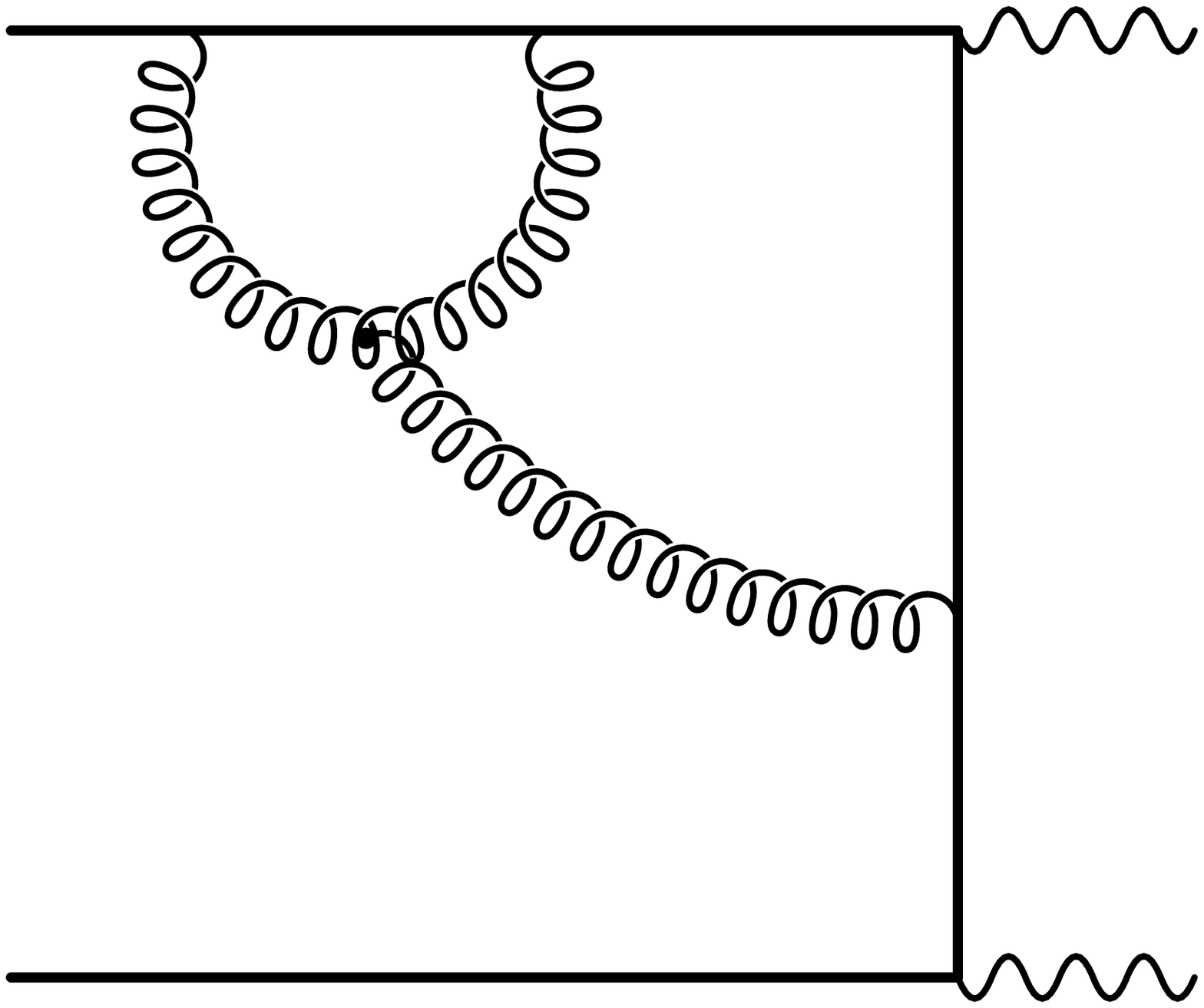,width=25mm}
\\[1mm]
 (a) & (b) & (c) 
\\[4mm]
\psfig{figure=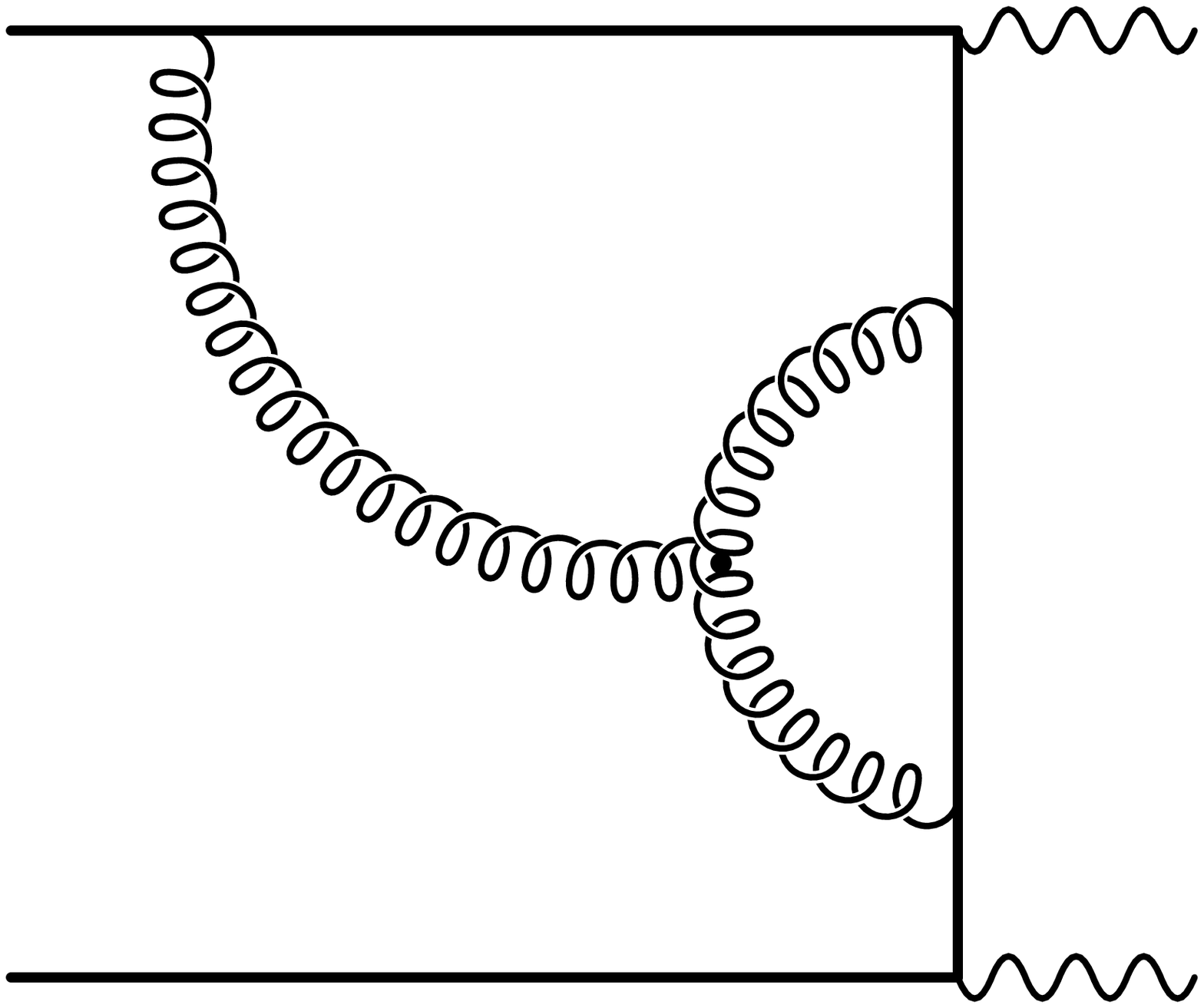,width=25mm}
&\hspace*{2mm}
\psfig{figure=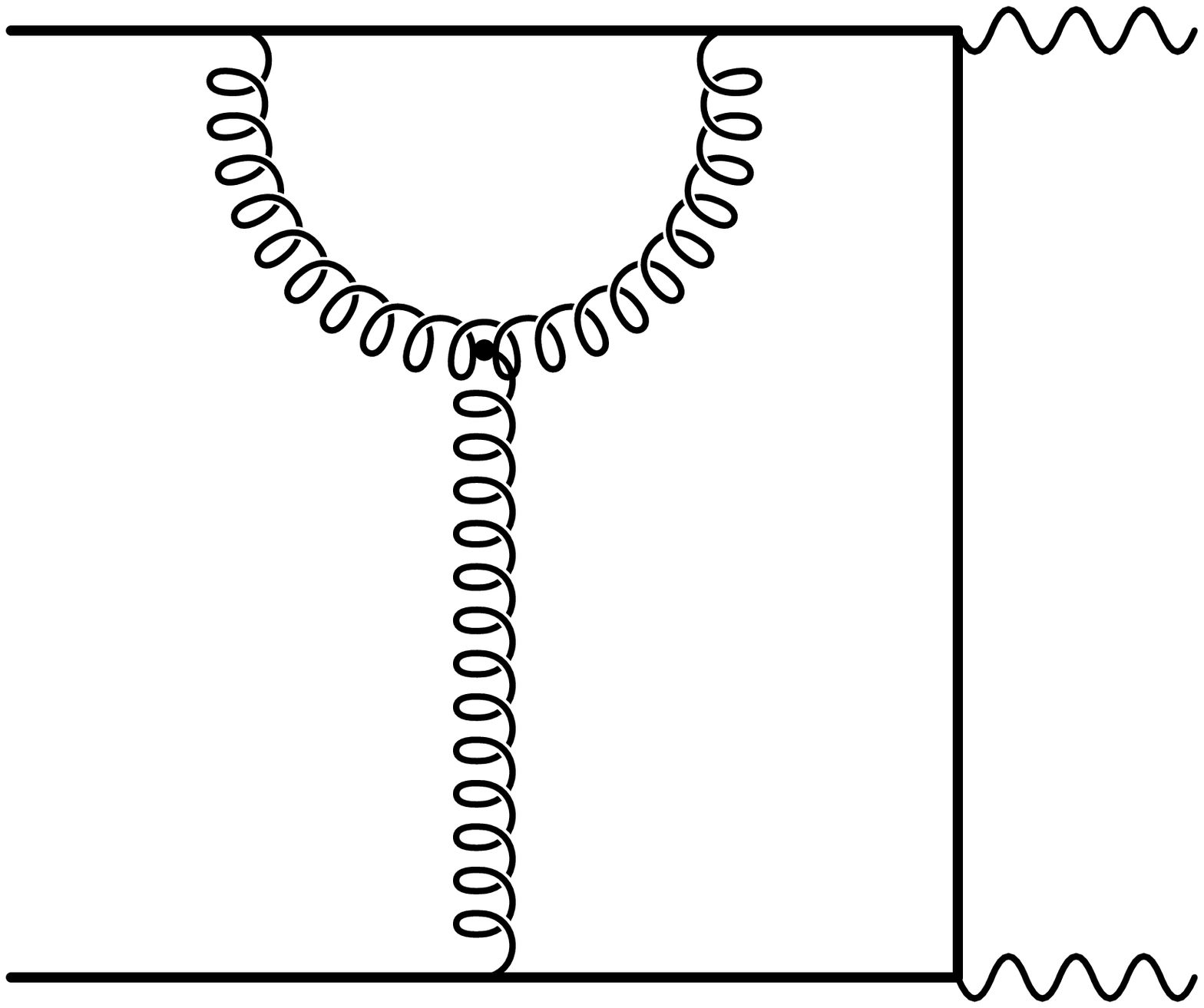,width=25mm}
&\hspace*{2mm}
\psfig{figure=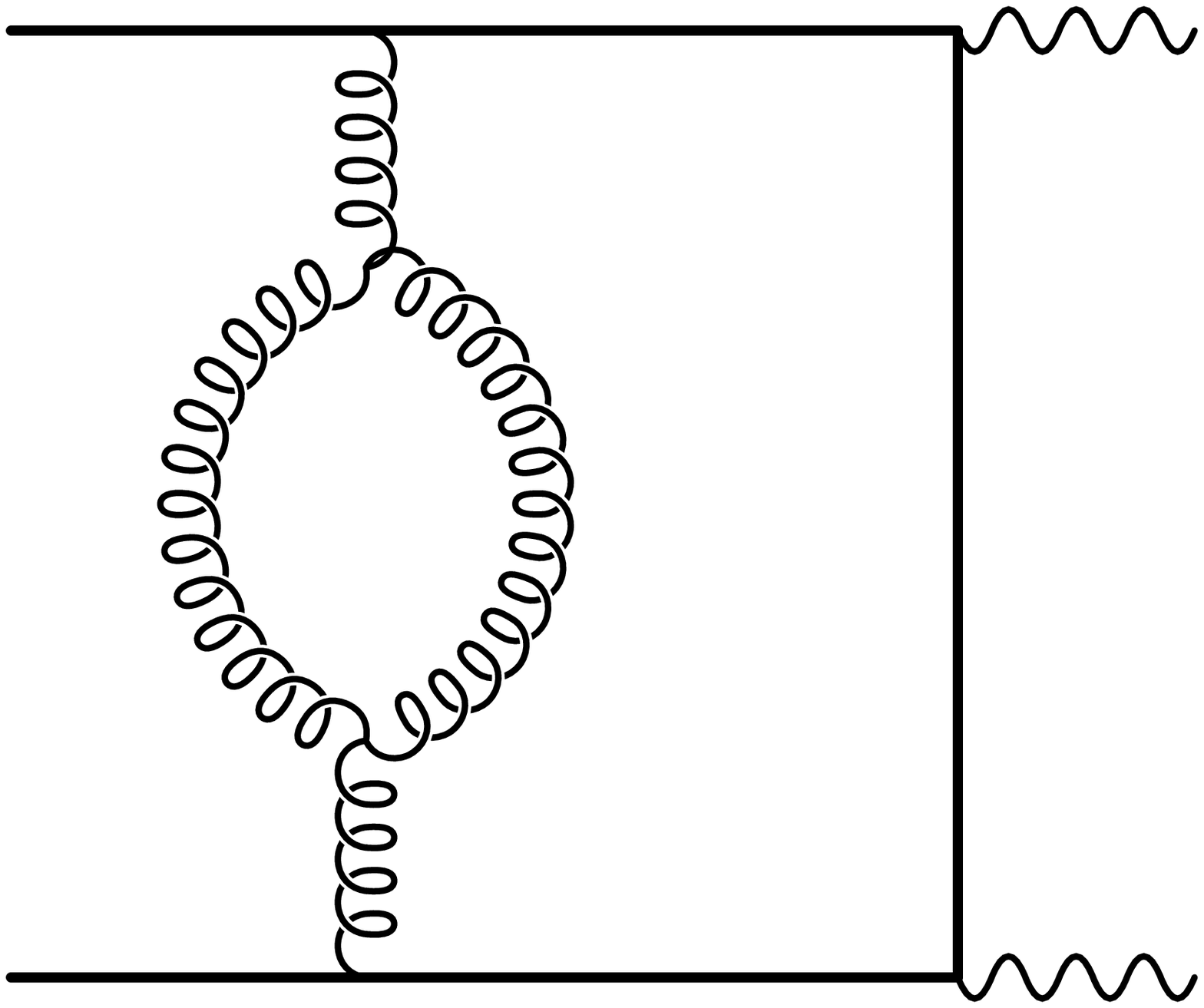,width=25mm}
\\[1mm]
 (d) & (e) & (f) 
\\[1mm]
\end{tabular}
\end{minipage}
\caption{Examples of non-abelian hard corrections to $\gamma\gamma\to 
t\overline t$ computed in this paper.  In addition, we
include  diagrams similar to (f) with gluon, Faddeev-Popov ghost and
fermion loops inserted in all one-loop corrections.}
\label{fig:na}
\end{figure}

To evaluate the hard renormalization factors we have used some of the
results obtained for the para-positronium decay
\cite{Czarnecki:1999gv,Czarnecki:1999ci}.  In addition, to find
$x_{NA}$ which is our main new result here, we had to evaluate the
non-abelian diagrams (see Fig.~\ref{fig:na}(a-e)) and massless
insertions in the one-loop diagrams (an example is shown in
Fig.~\ref{fig:na}(f)).  Here we briefly summarize those calculations.
Diagram \ref{fig:na}(a) is the simplest two-loop non-abelian
contribution.  It is both ultraviolet (UV) and infrared (IR) finite
and we evaluate it using Monte Carlo integration over six Feynman
parameters. (We use FORM \cite{form3} for symbolic manipulations and
Vegas \cite{Vegas} for numerical integrations.)  Diagram
\ref{fig:na}(b) is UV divergent.  To compute it we use a trick: assign
mass $M$ to the quark line (inside the loops) and expand the diagram
in the ratio $t=m^2/M^2$, treating $M$ as much larger than $m$.  At
the end we have to evaluate the sum of the resulting series at $t=1$.
It is sufficient to compute only a few first terms of the series in
$t$ if we change the variables, $t \to -4w/(1+w)^2$, expand in $w$,
and evaluate the new series at $w= 2\sqrt{2}-3\simeq -0.17$.

Diagrams (c) and (d) are also UV-divergent, with a subdivergence in
the non-abelian vertex correction and an overall divergence.  It is
useful to first remove the subdivergence by subtracting a product of
one-loop diagrams which can be evaluated analytically.  Next, the
remaining overall divergence is removed by subtracting a similar
two-loop amplitude which depends either only on the external quark
momentum (for (c)) or only on the spatial photon momentum (for (d)).
The resulting finite integrals are evaluated with Vegas, while the
subtracted divergent pieces can be computed either exactly (for (c))
or with the expansion used for the diagram (b).

The only IR-divergent non-trivial two-loop diagram is (e).  It also
contains a UV-subdivergence which we remove in the same way as in (c)
and (d).  The IR-divergence is subtracted by neglecting the loop
momentum in the ``t-channel'' quark propagator.  That subtraction can
be computed analytically, since it is simply a 2-loop vertex diagram
at threshold, of the type for which we developed a general algorithm
in an earlier study \cite{threshold}.  The finite difference is
evaluated with Vegas.

We did not include effects of the $\gamma\gamma\to gg$ scattering via
light quark boxes.  We have checked with a rough approximation that
this finite and gauge-invariant subset contributes only
insignificantly.

Using the result \eq{eq:rplusplus} it is easy to obtain the top
threshold production cross section for the photons with equal
helicities close to the threshold. In doing that, we follow the 
approach
described in Ref.~\cite{Melnikov:1998pr}. 
Let us first present a general formula for the resummed cross section
for two different choices of the photon helicities. We find:
\ba
&& R^{++}_S = \frac{24 \pi N_c Q_t^4}{m_t^2} C_h 
\nonumber \\
&& \times 
{\rm Im } \left \{ \left ( 1 - \frac{5}{6} \beta^2 \right )
G^{\rm sing} \left ( 0,0,E+i \Gamma_t \right ) \right \},
\ea
where $C_h$ is the hard renormalization factor and 
$G^{\rm sing}$ is the Green's function of the 
spin singlet state of the $t \bar t$ pair including  relativistic 
corrections and corrections to the Coulomb potential 
(see \cite{Melnikov:1998pr} for details).

As we already mentioned, there is also a $P$ wave contribution to the
$++$ cross section at NNLO.  It is suppressed by ${\cal O}(\beta^2)$
relative to the $S$ wave and working to NNLO we need to know it only
to leading order. Since for the photon helicities $+-$ the top pair is
also produced in the $P$ wave, there is a leading-order relation 
between 
$R^{++}_P$ and $R^{+-}$,
\ba
R^{+-} &=& \frac{4}{3} R^{++}_P, \\
R^{+-} &=& \frac{32 \pi Q_t^4 N_c}{m_t^4}
\partial_{\bf x} \partial_{\bf y} 
{\rm Im }\, G\left ({\bf x},{\bf y},E+i\Gamma_t \right )|_{{\bf 
x=0},{\bf y}=0}. 
\nonumber 
\ea

\begin{figure}[htb]
\vspace*{20mm}
\hspace*{-50mm}
\begin{minipage}{4.cm}
\begin{picture}(100,100)
\put (200,-10) {E [GeV]}
\put (70,25) {LO}
\put (190,75) {NLO}
\put (25,100) {$\mu=30$ GeV}
\put (95,100) {$60$ GeV}
\put (50,120) {NNLO}
\psfig{figure=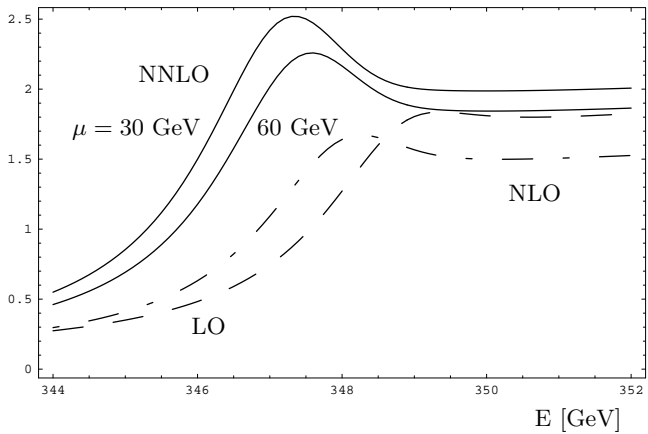,width=85mm}
\end{picture}
\end{minipage}
\vspace*{3mm}

\caption{$R_S^{++}$ at LO, NLO and NNLO. 
Top quark  pole mass $m_t = 175.05~{\rm GeV}$
width  $\Gamma_t = 1.43~{\rm GeV}$ are used.
The hard renormalization 
scale and the factorization scale are equal to the top quark mass
($\alpha_s(m_t) = 0.109$).
The soft renormalization scale is $60~{\rm GeV}$ ($\alpha_s(60~{\rm 
GeV})
=0.127$) but we also 
show the NNLO curve for the soft renormalization scale $30~{\rm GeV}$ 
to demonstrate large changes with the coupling constant scale 
variation.} 
\label{fig:plot}
\end{figure}

Numerically (see e.g. \cite{Penin:1998ik}) the $P$ wave contribution
is small; the corresponding values of $R^{+-}$ are $\sim 5 \times
10^{-2}$.  For this reason, we do not consider the $P$ wave
contribution in what follows.  In Fig.~\ref{fig:plot} we present the
LO, NLO and NNLO excitation curves for $R^{++}_S$ computed using the
pole mass of the top quark. As is clearly seen from this plot, the
NNLO corrections to the normalization of the cross section are quite
large: close to the peak they are about $20-30$\%. The position of the
peak of the cross section from which the mass of the top quark is to
be determined suffers from significant shifts when one goes from LO to
NNLO. All these features are quite similar to the known behavior of
$e^+e^- \to t \bar t$ at the threshold \cite{Hoang:2000yr}.

\begin{figure}[htb]
\vspace*{20mm}
\hspace*{-50mm}
\begin{minipage}{4.cm}
\begin{picture}(100,100)
\put (200,-10) {E [GeV]}
\put (35,120) {$\mu=30$ GeV}
\put (105,120) {$60$}
\put (50,140) {NNLO}
\psfig{figure=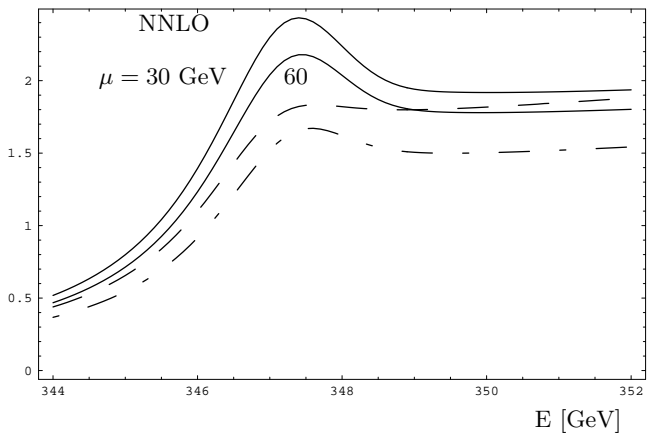,width=85mm}
\end{picture}
\end{minipage}
\vspace*{3mm}

\caption{
$R_S^{++}$ parameterized with the kinetic
mass $m_{\rm kin}(15~{\rm GeV}) = 173.10~{\rm GeV}$. Other 
parameters are the same as in Fig.~\ref{fig:plot}. 
Dashed, dashed-dotted and solid curves are LO, NLO and  NNLO
approximations, respectively.} 
\label{fig:plotKin}
\end{figure}

Let us now address these problems in turn. It was argued 
\cite{Beneke:1998rk,Hoang:1998nz}
in connection
with the threshold production of $t \bar t$ in $e^+e^-$ collisions
that significant shifts in the position of the peak are the
consequences of the fact that the pole mass scheme is unstable against
radiative corrections. The way out of this problem is to adopt a
different mass definition which will have such a stability. In
Fig.~\ref{fig:plotKin} we show the $S$-wave part of the cross
section, parameterized by the so-called kinetic mass
\cite{Bigi:1997si}.  We observe that the stability of the peak does
improve significantly, a behavior familiar from $e^+e^- \to t \bar t$
studies.

\begin{figure}[htb]
\vspace*{20mm}
\hspace*{-50mm}
\begin{minipage}{4.cm}
\begin{picture}(100,100)
\put (200,-10) {E [GeV]}
\put (20,20) {LO}
\put (60,40) {NLO}
\put (20,80) {NNLO, $\mu=60$ GeV}
\put (110,120) {NNLO, $\mu=30$ GeV}
\psfig{figure=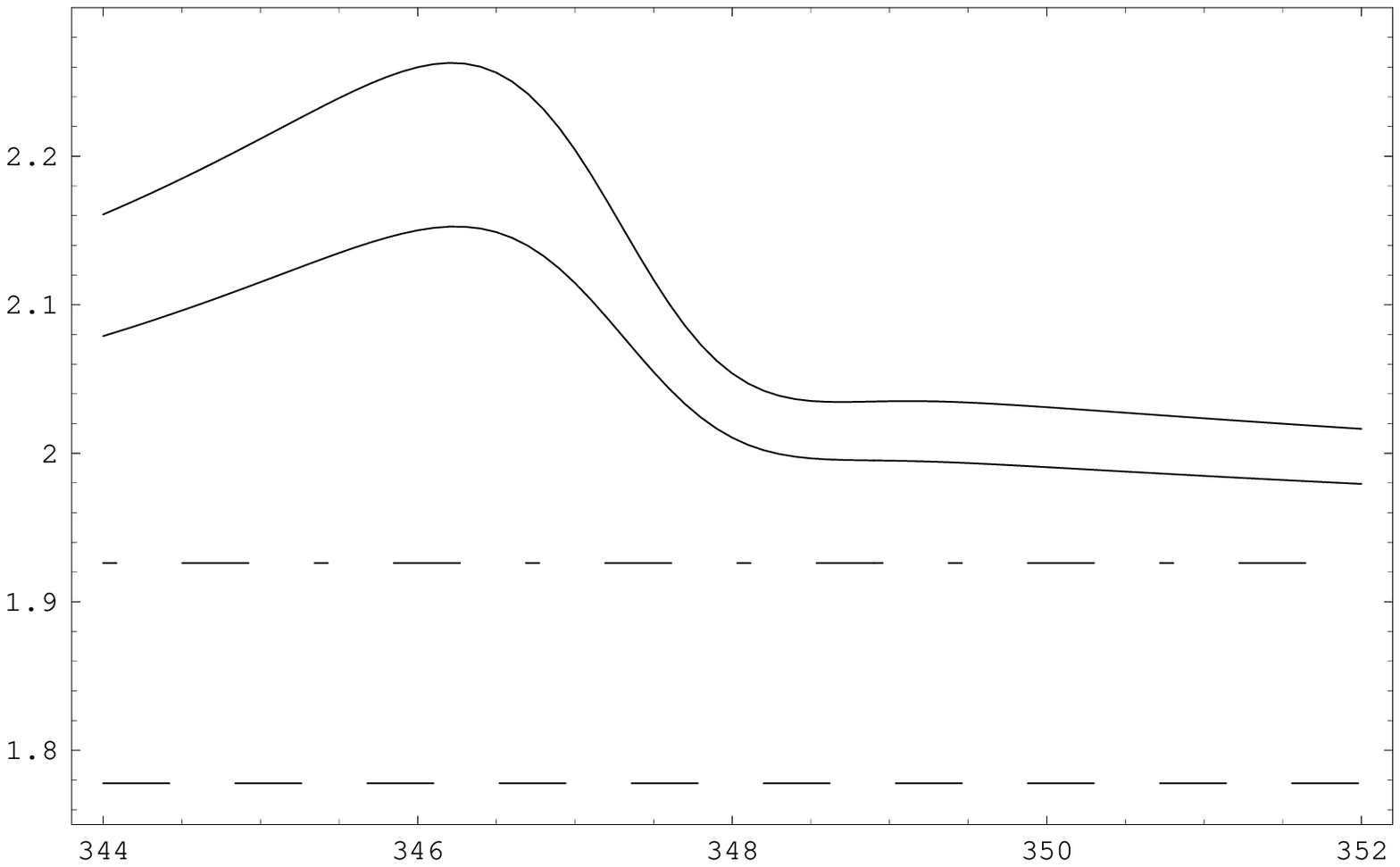,width=85mm}
\end{picture}
\end{minipage}
\vspace*{3mm}

\caption{
The ratio of $R^{++}_S$ and $R_e$ in the threshold region
at LO, NLO and NNLO (with two different soft renormalization
scales) parameterized by kinetic mass $m_{\rm kin}(15~{\rm GeV})$.
}
\label{fig:plot1kin}
\end{figure}

Let us now discuss the corrections to the peak height. Again, the 
situation
is similar to $e^+e^- \to t \bar t$ but the
corrections are larger. Since the height of the cross section is
controlled by the wave function at the origin, $\psi(0)$, of the $t
\bar t$ ground state, it is easy to understand what is happening by
looking at the $\ln \alpha_s$ enhanced corrections to $\psi(0)$
computed in \cite{Kniehl:1999mx}. Neglecting effects of the running of
the coupling constant, they read
\ba
\psi^2(0) & =& \psi_0^2(0) \left [ 1 - \alpha_s^2 \ln \alpha_s \left ( 
7.55 - 1.19~S(S+1) \right)
\right.
\nonumber \\ 
&& \left. + \alpha_s^3 \ln^2 \alpha_s \left ( -9.48 + 0.99~S(S+1) 
\right)
\right ].
\label{eq12}
\ea
The ${\cal O}(\alpha_s^3)$ (N$^3$LO) leading logarithmic effects are
included in this formula. From Eq.~(\ref{eq12}) one sees that there
is, effectively, a compensation between spin-dependent and
spin-independent terms in the correction to the wave function at the
origin; since the top quarks are produced in the spin-singlet state in
$\gamma \gamma \to t \bar t$, the corrections in this case are
expected to be larger than in $e^+e^- \to t \bar t$. Note also that in
both cases the $\alpha_s^3 \ln \alpha_s$ corrections are negative, so
that the NNLO curves shown in Fig.~\ref{fig:plot} are expected to be
pushed down in N$^3$LO, which clearly illustrates the sign-alternating
nature of the perturbative series. The success of the resummation
program \cite{Hoang:2001mm} for $e^+e^- \to t \bar t$ is related to
this property of the series; for this reason one can expect a
significant improvement in the stability of the height of the cross
section also in $\gamma \gamma \to t \bar t$ once the resummation
program is carried out.

It is also interesting to consider the ratio of the top threshold
production cross sections in $e^+e^-$ and $\gamma \gamma$ collisions,
plotted in Fig.~\ref{fig:plot1kin}.  We see that the ratio is energy
independent at LO and NLO, since in these orders the spin of the $t
\bar t$ pair decouples. However, at the NNLO the non-relativistic
Hamiltonian depends on the spin of the $t \bar t$ pair 
and the ratio of the cross-sections
becomes energy dependent. This results in large corrections to the
ratio in the vicinity of the peak.  Away from the peak where
the bound state dynamics is not very important and the ratio is,
essentially, given by the hard renormalization factors of the
production currents, the convergence of the perturbative series for
the ratio is quite good.

The NNLO corrections to the top pair threshold production in
$\gamma\gamma$ collisions, which we have completed by evaluating the
non-abelian hard renormalization factor, turn out to be large.
This is similar to the process $e^+e^-\to t\bar t$.  In both cases we
see that the position of the cross section peak is stabilized when a
short-distance quark mass definition is used, but the height of the
peak is changed significantly by the NNLO corrections. We have argued 
that the structure of the NNLO corrections to $\gamma \gamma \to t \bar 
t$ 
at the threshold is rather similar to that of $e^+e^- \to t \bar t$; 
in particular the normalization of the cross section is determined 
by sign-alternating series. This suggests that if the renormalization 
group improvement, applied to $e^+e^- \to t \bar t$ in 
\cite{Hoang:2001mm}
is applied to $\gamma \gamma \to t \bar t$, one can expect significant 
improvement in the stability of the normalization of the production 
cross section. For this reason,   in the
future, it would be interesting to resum the  logarithmic
corrections $\order{\ln(\beta)}$ for $\gamma\gamma \to t\bar t$,
along the lines of such an analysis for the $e^+e^-$ annihilation
\cite{Hoang:2001mm}.  We hope that a combined analysis of both
processes will shed new light on the behavior of higher-order
corrections to the threshold processes and allow to reach better 
theoretical precision.

A.C. thanks A. S. Yelkhovsky for helpful discussions.  This research
was supported in part by the Natural Sciences and Engineering Research
Council of Canada and by the DOE under grant number
DE-AC03-76SF00515.


\end{document}